\documentclass[
  aps,
  prl,
  twocolumn,
  superscriptaddress,
  nofootinbib,
  floatfix, showpacs
]{revtex4-2}

\usepackage{graphicx}
\usepackage{dcolumn}
\usepackage{bm}

\usepackage{amsmath}
\usepackage{amssymb}
\usepackage{siunitx} 
\usepackage{xcolor} 
\usepackage[normalem]{ulem}

\usepackage[
  colorlinks=true,
  linkcolor=black,
  citecolor=blue!50!black,
  urlcolor=blue!50!black
]{hyperref}

\begin{document}

\title{Resonant neutrino flavor conversion within dark matter spikes}

\author{P. S. Bhupal Dev}
\email{bdev@wustl.edu}
\affiliation{Department of Physics and McDonnell Center for the Space Sciences,
Washington University, One Brookings Drive, St. Louis, MO 63130, USA}

\author{Elisa Gaido}
\email{elisa.gaido@tum.de}
\affiliation{Technical University of Munich, TUM School of Natural Sciences, Physics Department, 85748 Garching, Germany}

\author{Alejandro Ibarra}
\email{ibarra@tum.de}
\affiliation{Technical University of Munich, TUM School of Natural Sciences, Physics Department, 85748 Garching, Germany}

\author{Yago Porto}
\email{yago.porto@tum.de}
\affiliation{Technical University of Munich, TUM School of Natural Sciences, Physics Department, 85748 Garching, Germany}

\begin{abstract}
We investigate how neutrino--dark matter (DM) interactions modify the flavor composition of high-energy neutrinos from active galactic nuclei (AGNs). Coherent forward scattering in a DM spike around the central supermassive black hole can generate a flavor-dependent potential, inducing flavor conversion as the neutrinos escape. We calculate the flavor composition at Earth for pion-decay and muon-damped sources and find substantial departures from vacuum-oscillation expectations. At neutrino energies around $100\,\mathrm{TeV}$, the matter potential begins to compete with vacuum oscillations when the coupling-weighted net DM number density reaches $|\epsilon_\alpha(n_\chi-n_{\bar\chi})|\sim10^{18}\,\mathrm{cm}^{-3}$. We compare our predictions with IceCube flavor triangle measurements under the illustrative assumption that the diffuse flux originates from AGNs with common neutrino-production and DM-spike properties. Under this assumption, when the DM-induced potential acts on the muon flavor and dominates the vacuum terms at production, the $p\gamma$ muon-damped predictions lie outside the 95\% MESE contour. High-energy neutrino flavor measurements can therefore provide a novel probe of neutrino-DM interactions in astrophysical environments.
\end{abstract}

\maketitle

\section{Introduction}
\label{sec:Intro} 

High-energy astrophysical neutrinos provide a new window on the most energetic environments in the Universe and a unique probe of particle physics over cosmological baselines. Since IceCube established the existence of an astrophysical neutrino flux~\cite{IceCube:2013cdw,IceCube:2013low}, growing evidence suggests that active galactic nuclei (AGNs) may contribute significantly to it~\cite{Halzen:2023usr, Jain:2026jdh}. This includes the multimessenger association of IceCube-170922A with TXS~0506+056 and the archival neutrino flare from its direction~\cite{IceCube:2018cha, IceCube:2018dnn}, as well as evidence for neutrino emission from NGC~1068~\cite{IceCube:2022der,IceCube:2025lev}. More recently, an IceCube population analysis found an excess associated with X-ray-bright Seyfert galaxies, further supporting neutrino production in AGN cores~\cite{IceCube:2025gdd}.

If, as current evidence suggests, AGNs contribute substantially to the diffuse astrophysical neutrino flux, its flavor composition can reveal how neutrinos are produced and how they propagate from these sources~\cite{IceCube:2025ole}. The source flavor composition is determined by the dominant neutrino-production and cooling processes. In the standard pion-decay scenario, the source flavor ratio $(\nu_e:\nu_\mu:\nu_\tau)_{\rm S}\simeq(1/3:2/3:0)$ evolves through vacuum oscillations into a composition close to $(1/3:1/3:1/3)_{\oplus}$ on Earth~\cite{Learned:1994wg}, whereas the muon-damped pion decay scenario, $(0:1:0)_{\rm S}$, leads approximately to $(0.2:0.4:0.4)_{\oplus}$~\cite{Rachen:1998fd,Beacom:2003nh,Kashti:2005qa,Hummer:2010ai,Dev:2023znd}. New interactions at the source or detector, or during propagation, can shift these standard expectations~\cite{Arguelles:2015dca,Arguelles:2019tum,Ahlers:2020miq,Song:2020nfh,Arguelles:2024cjj,Fong:2024msb,Dev:2024yrg,Telalovic:2025xor,Bustamante:2026zst,Verma:2026wrs}. In particular, if neutrinos interact with dark matter (DM), coherent forward scattering generates a matter potential, in close analogy with the standard MSW effect in ordinary matter~\cite{Wolfenstein:1977ue,Mikheyev:1985zog}. The presence of a DM density spike~\cite{Gondolo:1999ef} at the astrophysical source can significantly enhance this `dark' matter effect for neutrinos.  For the Milky Way, we assume that no steep DM spike survives around Sgr~A$^*$, since stellar heating and other dynamical processes may substantially soften an initially formed spike~\cite{Ullio:2001fb,Merritt:2003qk,Bertone:2005hw}, as supported by constraints from recent observations of stellar orbits~\cite{Lacroix:2018zmg,Shen:2023kkm,Ou:2023adg,GRAVITY:2024tth} and  DM indirect detection searches~\cite{Chattopadhyay:2026kbm}. By contrast, we assume that a steep DM spike survives around the central supermassive black hole (SMBH) of the source AGN~\cite{Wang:2021jic,Cline:2022qld,Ferrer:2022kei,DeMarchi:2024riu,Meighen-Berger:2025hrq}, so that high-energy neutrinos produced near the SMBH traverse an exceptionally dense DM environment before escaping, making AGNs natural laboratories in which to search for neutrino--DM interactions through their effects on flavor oscillations. Neutrino flavor conversion in smooth Galactic DM halos has been studied in Refs.~\cite{deSalas:2016svi, Choi:2019zxy}. When translated into our framework, an unspiked NFW profile would require couplings approximately $10^{9}$--$10^{10}$ times larger than in our spike scenario to compensate for the lower DM density and generate comparable matter potentials.  

In this Letter, we consider high-energy neutrinos produced near the central SMBHs of AGNs~\cite{Inoue:2019yfs,Murase:2022dog,Padovani:2024ibi} and solve their complete three-flavor evolution through the surrounding DM spikes. We determine how neutrino energy and the magnitude, sign, and flavor structure of the DM-induced potential affect the flavor composition at Earth for pion-decay and muon-damped sources. We show that DM-induced flavor conversion, including nonadiabatic transitions, can produce substantial departures from the standard vacuum-oscillation predictions. These differences survive phase averaging over astrophysical baselines. We compare the resulting flavor ratios with current IceCube flavor measurements~\cite{IceCube:2025ole} and IceCube-Gen2 projections~\cite{IceCube-Gen2:2023rds}. Our calculation focuses on coherent forward scattering, neglecting hard scattering and the associated attenuation and regeneration of the flux. A recent treatment including these effects within a specific particle-physics model is given in Ref.~\cite{Abbaslu:2026row}.


\section{Dark matter effects in neutrino oscillations}
\label{sec:DM-effects}

For concreteness, we assume that DM is a Dirac fermion $\chi$ interacting with neutrinos through the flavor-dependent effective vector interaction
\begin{align}
\mathcal{L}_{\rm int}
=
-2\sqrt{2}\,G_F
\sum_{\alpha=e,\mu,\tau}
\epsilon_\alpha
\left(\overline{\nu}_{L\alpha}\gamma^\rho\nu_{L\alpha}\right)
\left(\overline{\chi}\gamma_\rho\chi\right).
\label{eq:interaction}
\end{align}
Here, $\alpha=e,\mu,\tau$ denotes the neutrino flavor, and the dimensionless parameters $\epsilon_\alpha$ characterize the corresponding interaction strengths normalized to the Fermi constant $G_F$. This form is analogous to the Wolfenstein parametrization of the nonstandard neutrino interactions~\cite{Wolfenstein:1977ue}. Their values can be either larger or smaller than unity, depending on the mediator mass and its couplings to neutrinos and DM. The overall minus sign is chosen so that $\epsilon_\alpha>0$ corresponds to a positive matter potential for neutrinos propagating through a particle-dominated DM background.

Consider a neutrino beam with energy $E$ propagating through a nonrelativistic medium of Dirac DM particles and antiparticles with number densities $n_\chi$ and $n_{\bar{\chi}}$, respectively. The total DM mass density and the net DM number density are
\begin{equation}
\rho_\chi=m_\chi\left(n_\chi+n_{\bar{\chi}}\right),
\qquad
\Delta n_\chi=n_\chi-n_{\bar{\chi}}.
\end{equation}
The effective Hamiltonian governing neutrino propagation through this medium is
\begin{equation}
H_{\rm eff}
=
\frac{1}{2E}\,
U
\begin{pmatrix}
0 & 0 & 0\\
0 & \Delta m_{21}^2 & 0\\
0 & 0 & \Delta m_{31}^2
\end{pmatrix}
U^\dagger
+
\begin{pmatrix}
V_e & 0 & 0\\
0 & V_\mu & 0\\
0 & 0 & V_\tau
\end{pmatrix}.
\label{eqn:Heff_mu}
\end{equation}
Here, $U$ is the leptonic mixing matrix and $\Delta m_{ij}^2=m_i^2-m_j^2$, where $m_i$ are the neutrino masses. An irrelevant contribution proportional to the identity has been subtracted from the vacuum Hamiltonian. The expression therefore applies to both mass orderings, with the sign of $\Delta m_{31}^2$ distinguishing normal from inverted ordering. The flavor-dependent potential generated by coherent forward scattering on the DM background for neutrinos is
\begin{align}
V_\alpha
&=
2\sqrt{2}\,G_F\epsilon_\alpha\Delta n_\chi
\nonumber\\
&\simeq
\SI{2.5e-19}{eV}
\left(
\frac{\epsilon_\alpha\Delta n_\chi}
{10^{18}\,\mathrm{cm}^{-3}}
\right).
\label{eqn:muon_potential}
\end{align}
For antineutrinos, the Hamiltonian is obtained by replacing $U\rightarrow U^*$ and $V_\alpha\rightarrow -V_\alpha$ in Eq.~\eqref{eqn:Heff_mu}. The factor of two relative to the conventional MSW potential follows from the use of the full DM vector current, for which $\langle\overline{\chi}\gamma^0\chi\rangle=\Delta n_\chi$, whereas the expectation value of the chiral electron current in an unpolarized medium is $n_e/2$. Consequently, a particle--antiparticle symmetric DM background produces no matter potential through the interaction in Eq.~\eqref{eq:interaction}.

For neutrinos propagating in vacuum, the central values from NuFIT~6.0~\cite{Esteban:2024eli} give characteristic matrix elements in the range
\begin{equation}
\left|(H_{\rm vac})_{\alpha\beta}\right|
\sim
\left(3.9\times10^{-19}
      -7.1\times10^{-18}\right)\,\mathrm{eV}
\left(\frac{100\,\mathrm{TeV}}{E}\right),
\label{eq:vacuum_scale}
\end{equation}
for normal ordering, and similar values for inverted ordering. Comparing this scale with the DM-induced potential in Eq.~\eqref{eqn:muon_potential} shows that, at $E\sim100\,\mathrm{TeV}$, matter effects can begin to compete with vacuum oscillations when the coupling-weighted net DM number density reaches $\left|\epsilon_\alpha\Delta n_\chi\right|\sim10^{18}\,\mathrm{cm}^{-3}$, although the precise onset of flavor conversion depends on the sign and flavor structure of the potential. As we show in the next section, the large net DM densities attainable in the spikes surrounding the central SMBHs of AGNs can bring $\left|\epsilon_\alpha\Delta n_\chi\right|$ into this regime.

\section{Dark matter spikes in active galactic nuclei}

We model the DM distribution surrounding the central SMBH of an AGN as in Ref.~\cite{Akita:2025dhg}, matching the central spike to an outer NFW profile~\cite{Navarro:1995iw},
\begin{equation}
\rho_{\rm NFW}(r)
=
\rho_s
\left(\frac{r}{r_s}\right)^{-1}
\left(1+\frac{r}{r_s}\right)^{-2},
\label{eq:nfw_profile}
\end{equation}
where $r_s$ and $\rho_s$ are the scale radius and scale density, respectively. The adiabatic growth of the central SMBH steepens the original $r^{-1}$ cusp into a spike proportional to $r^{-7/3}$~\cite{Gondolo:1999ef}. In the nonannihilating limit, we adopt
\begin{equation}
\rho_{\rm sp}(r)
=
\rho_R
\left(1-\frac{4R_S}{r}\right)^3
\left(\frac{R_{\rm sp}}{r}\right)^{7/3},
\label{eq:spike_profile}
\end{equation}
for $4R_S\leq r\leq R_{\rm sp}$, where $R_S=2GM_{\rm BH}$ is the Schwarzschild radius of the SMBH in natural units, $M_{\rm BH}$ is the SMBH mass, and $R_{\rm sp}=0.2GM_{\rm BH}/v_0^2$ is the spike radius, with $v_0$ being the bulge velocity dispersion. We set the density to zero for $r<4R_S$ and match the spike continuously to the NFW profile at $R_{\rm sp}$. This fixes its normalization to $\rho_R=\rho_{\rm NFW}(R_{\rm sp})\left(1-4R_S/R_{\rm sp}\right)^{-3}$.

\begin{figure}[!t]
    \centering
    \includegraphics[width=\columnwidth]{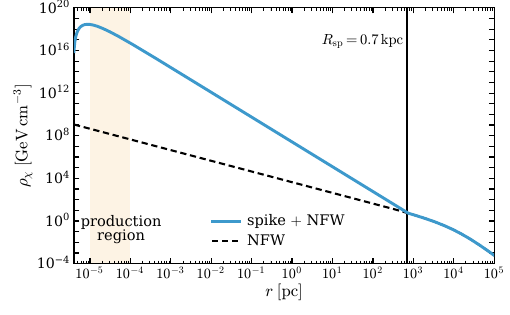}
    \caption{DM density profile for the NGC~1068 benchmark. The solid blue curve shows the adopted spike matched to the outer NFW halo, while the dashed black curve shows the unmodified NFW profile. The shaded band marks the neutrino-production region, and the vertical line denotes $R_{\rm sp}=\SI{0.7}{kpc}$.}
    \label{fig:dm_spike_profile}
\end{figure}

Following the NGC~1068 benchmark, we take $M_{\rm BH}=10^7M_\odot$, $r_s=\SI{13}{kpc}$, $\rho_s=\SI{0.35}{GeV/cm^3}$, and $R_{\rm sp}=\SI{0.7}{kpc}$. Coronal models and multimessenger studies of NGC~1068 motivate neutrino production at radii of order $10$--$100R_S$, corresponding approximately to $10^{-5}$--$10^{-4}\,\mathrm{pc}$ for this SMBH mass~\cite{Inoue:2019yfs,Murase:2022dog,Padovani:2024ibi}. Some analyses of photohadronic production favor even more compact emission regions~\cite{Das:2024vug}. We denote the neutrino-production radius by $r_{\rm p}$ and, for concreteness, fix $r_{\rm p}=\SI{e-5}{pc}$. As shown in Fig.~\ref{fig:dm_spike_profile}, this radius lies close to the maximum of the spike density at $r\simeq\SI{9e-6}{pc}$.

For a maximally asymmetric background composed entirely of either DM particles or DM antiparticles~\cite{Petraki:2013wwa,Zurek:2013wia}, the coupling-weighted net DM number density is
$|\epsilon_\alpha\Delta n_\chi(r)|=|\epsilon_\alpha|\rho_\chi(r)/m_\chi$. At the production radius, $\rho_\chi(r_{\rm p})\simeq\SI{3e18}{GeV/cm^3}$ and hence
$|\epsilon_\alpha\Delta n_\chi(r_{\rm p})|\simeq3\times10^{18}|\epsilon_\alpha|(m_\chi/\mathrm{GeV})^{-1}\,\mathrm{cm}^{-3}$. Thus, comparing Eqs.~\eqref{eqn:muon_potential} and \eqref{eq:vacuum_scale}, we find that for $|\epsilon_\alpha|(m_\chi/\mathrm{GeV})^{-1}\gtrsim1$, the DM-induced potential can compete with the vacuum terms at energies around $\SI{100}{TeV}$.

\section{Neutrino propagation through dark matter spikes}
\label{sec:spike_propagation}

For a background composed entirely of DM particles, $\Delta n_\chi(r)=\rho_\chi(r)/m_\chi$. Eq.~\eqref{eqn:muon_potential} then gives the radial potential for neutrinos,
\begin{align}
V_\alpha(r)
&\simeq
\SI{7.5e-19}{eV}\,
\epsilon_\alpha
\left(\frac{m_\chi}{\mathrm{GeV}}\right)^{-1}
\nonumber\\[-2pt]
&\quad\times
\left(
\frac{\rho_\chi(r)}
{\SI{3e18}{GeV/cm^3}}
\right).
\label{eq:radial_potential}
\end{align}
With the production radius fixed at $r_{\rm p}=\SI{e-5}{pc}$, the propagation depends on the DM parameters through the combinations $\epsilon_\alpha(m_\chi/\mathrm{GeV})^{-1}$.

For a neutrino produced with flavor $\beta$, we solve the complete radial Schr\"{o}dinger equation,
\begin{equation}
i\frac{d}{dr}|\nu_\beta(r)\rangle
=
H_{\rm eff}(r,E)|\nu_\beta(r)\rangle,
\label{eq:radial_evolution}
\end{equation}
using the Hamiltonian in Eq.~\eqref{eqn:Heff_mu} and the potential in Eq.~\eqref{eq:radial_potential}. Let $|\nu_\beta^{\rm sp}\rangle$ denote the state obtained by evolving an initial flavor state $|\nu_\beta\rangle$ through the spike. Once the neutrino leaves the spike, the astrophysical source-to-Earth baseline averages the relative phases between the vacuum mass eigenstates. The resulting transition probability is
\begin{equation}
P_{\beta\alpha}^{\nu,\oplus}
=
\sum_i
\left|U_{\alpha i}\right|^2
\left|\langle\nu_i|\nu_\beta^{\rm sp}\rangle\right|^2.
\label{eq:propagation_probability}
\end{equation}
The numerical solution includes both adiabatic flavor conversion and nonadiabatic transitions. Appendix~\ref{app:adiabaticity} examines the adiabaticity of the evolution and identifies where nonadiabatic effects become relevant.

\begin{figure*}[t]
    \centering
    \includegraphics[width=\textwidth]{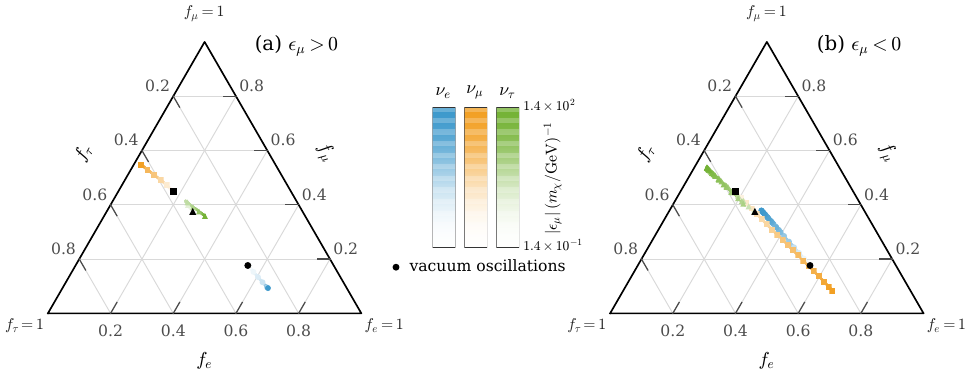}
    \caption{Flavor composition at Earth for initially pure $\nu_e$ (circles), $\nu_\mu$ (squares), and $\nu_\tau$ (triangles) fluxes produced with energy $E=\SI{100}{TeV}$. We assume normal ordering and a DM interaction acting only on the muon flavor. The complete three-flavor evolution through the spike is calculated numerically, including nonadiabatic transitions, after which the vacuum propagation to Earth is phase averaged. For a particle background, the left and right panels correspond to $\epsilon_\mu>0$ and $\epsilon_\mu<0$, respectively. Along each trajectory, the color intensity indicates increasing $|\epsilon_\mu|\left(m_\chi/\mathrm{GeV}\right)^{-1}$ from $1.4\times10^{-1}$ to $1.4\times10^{2}$. The black markers indicate the vacuum-oscillation limit.}
    \label{fig:example_muon}
\end{figure*}

Fig.~\ref{fig:example_muon} illustrates the resulting flavor evolution for initially pure $\nu_e$, $\nu_\mu$, and $\nu_\tau$ fluxes with energy $E=\SI{100}{TeV}$, assuming normal ordering and a DM interaction restricted to the muon flavor ($\epsilon_e=\epsilon_\tau=0$). Each colored point is obtained from the complete numerical evolution for a different value of $\epsilon_\mu\left(m_\chi/\mathrm{GeV}\right)^{-1}$. The left and right panels correspond to positive and negative values of this combination, respectively, for a particle background. The trajectories span
$1.4\times10^{-1}\leq|\epsilon_\mu|\left(m_\chi/\mathrm{GeV}\right)^{-1}\leq1.4\times10^{2}$. Near the lower end of this interval, the matter contribution is negligible and the flavor compositions approach the vacuum expectations indicated by the black markers. Departures become appreciable for values of order unity, when the DM contribution begins to compete with the vacuum Hamiltonian. Toward the upper end of the interval, the evolution approaches the matter-dominated limit and the flavor compositions effectively saturate. If the sign of $\epsilon_\mu$ is known, the difference between the two panels can distinguish a particle-dominated from an antiparticle-dominated DM background.

\section{Neutrino production and flavor evolution in active galactic nuclei}
\label{sec:agn-flavor}

Within AGNs, accelerated protons can produce high-energy neutrinos through $p\gamma$ interactions with ambient radiation or through $pp$ collisions with gas in winds and tori. Possible production sites extend from the accretion disk and corona near the central SMBH to the broad-line region, dust torus, and circumnuclear gas, and may also include the jet. Here we adopt the compact NGC~1068 production benchmark, $r_{\rm p}=\SI{e-5}{pc}$, motivated by coronal models and multimessenger studies~\cite{Inoue:2019yfs,Murase:2022dog,Padovani:2024ibi}.

The two source compositions introduced above can be understood from the charged-pion decay chain. For an idealized $p\gamma$ source dominated by $\pi^+$ production, $p+\gamma\rightarrow\Delta^+\rightarrow n+\pi^+$~\cite{Hummer:2010vx}, followed by $\pi^+\rightarrow\mu^++\nu_\mu$ and $\mu^+\rightarrow e^++\nu_e+\bar{\nu}_\mu$. The final state therefore contains one $\nu_e$, one $\nu_\mu$, and one $\bar{\nu}_\mu$, yielding the neutrino-plus-antineutrino source composition $\left(f_e^{\rm S},f_\mu^{\rm S},f_\tau^{\rm S}\right)=(1/3,2/3,0)$. If the secondary muon loses most of its energy before decaying, only the prompt $\nu_\mu$ from the $\pi^+$ decay contributes at high energies. This is the muon-damped regime, for which $\left(f_e^{\rm S},f_\mu^{\rm S},f_\tau^{\rm S}\right)=(0,1,0)$~\cite{Hummer:2010ai}. By contrast, $pp$ interactions produce approximately equal numbers of $\pi^+$ and $\pi^-$, while yielding the same flavor-summed source compositions in the idealized pion-decay and muon-damped limits~\cite{Biehl:2016psj}.

Because the matter potential has the opposite sign for antineutrinos, we propagate the neutrino and antineutrino components separately. The flavor fraction measured at Earth is
\begin{equation}
f_{\alpha,\oplus}
=
\sum_\beta
\left[
f_{\nu_\beta}^{\rm S}P_{\beta\alpha}^{\nu,\oplus}
+
f_{\bar{\nu}_\beta}^{\rm S}P_{\beta\alpha}^{\bar{\nu},\oplus}
\right],
\label{eq:combined_flavor}
\end{equation}
where $f_{\nu_\beta}^{\rm S}$ and $f_{\bar{\nu}_\beta}^{\rm S}$ are the charge-resolved source fractions, normalized such that $\sum_\beta\left(f_{\nu_\beta}^{\rm S}+f_{\bar{\nu}_\beta}^{\rm S}\right)=1$. For antineutrinos, we repeat the evolution and phase averaging used for Eq.~\eqref{eq:propagation_probability}, replacing $U$ by $U^*$ and $V_\alpha$ by $-V_\alpha$ in the Hamiltonian of Eq.~\eqref{eqn:Heff_mu}.

\begin{figure*}[t]
    \centering
    \includegraphics[width=\textwidth]{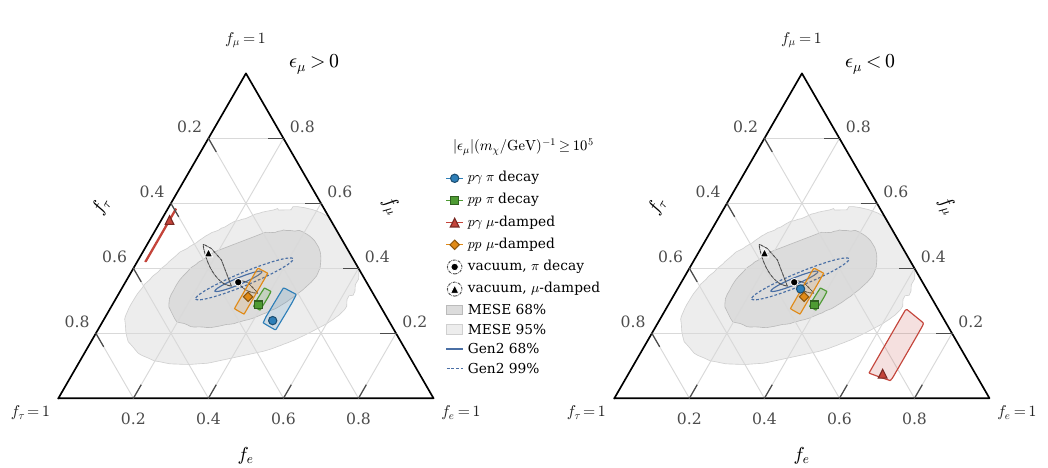}
    \caption{Flavor compositions at Earth for $p\gamma$ and $pp$ pion-decay and muon-damped sources with a DM interaction restricted to the muon flavor. The left and right panels correspond to $\epsilon_\mu>0$ and $\epsilon_\mu<0$ for a particle-only DM background. Colored markers show the central predictions in the regime $|\epsilon_\mu|(m_\chi/\mathrm{GeV})^{-1}\geq10^5$, where the DM-induced potential dominates the vacuum terms at production and the flavor compositions remain approximately constant across the MESE energy range; the surrounding regions are envelopes obtained by independently varying the oscillation parameters over their marginal $3\sigma$ ranges. Black markers and outlines show the corresponding vacuum predictions. The complete numerical evolution through the spike is used throughout. Gray regions denote the 68\% and 95\% C.L. MESE flavor contours~\cite{IceCube:2025ole}; blue solid and dashed curves show the illustrative 68\% and 99\% C.L. IceCube-Gen2 projections~\cite{IceCube-Gen2:2023rds}, respectively, assuming a pion-decay flavor composition as their true signal.}
\label{fig:comparison_icecube_2}
\end{figure*}

\section{Dark matter effects in IceCube flavor measurements}
\label{sec:icecube}

IceCube has identified neutrino emission from AGNs~\cite{IceCube:2018cha, IceCube:2022der}, yet their individual flavor compositions remain unmeasured, even in recent point-source searches combining track and cascade data~\cite{IceCube:2025lev}. By contrast, IceCube measures the flavor composition of the diffuse astrophysical flux. Tracks arise mainly from $\nu_\mu$ charged-current interactions, cascades receive contributions from $\nu_e$ and $\nu_\tau$ charged-current interactions and neutral-current interactions of all flavors, and double cascades provide additional sensitivity to $\nu_\tau$. The IceCube Medium-Energy Starting Events (MESE) analysis uses these event classes to constrain the diffuse flavor composition over \SI{5}{TeV}--\SI{10}{PeV}, reporting contours for a flavor ratio assumed to be independent of energy~\cite{IceCube:2025ole}.

To compare our AGN predictions with the diffuse measurement, we assume that AGNs dominate the flux~\cite{Jain:2026jdh} and share sufficiently similar neutrino-production conditions and DM spikes that the diffuse flavor composition can be approximated by that of a representative AGN. We use the $p\gamma$ and $pp$ pion-decay and muon-damped benchmarks introduced above, propagating their neutrino and antineutrino components separately as in Eq.~\eqref{eq:combined_flavor}.

For the adopted spike and an interaction restricted to the muon flavor, $|\epsilon_\mu|(m_\chi/\mathrm{GeV})^{-1}\lesssim10^{-1}$ gives flavor compositions close to the vacuum-oscillation limit across the MESE energy range, whereas $|\epsilon_\mu|(m_\chi/\mathrm{GeV})^{-1}\gtrsim10^5$ gives compositions close to those expected when the DM-induced potential dominates the vacuum terms at production and propagation through the spike is approximately adiabatic. Fig.~\ref{fig:comparison_icecube_2} compares these limiting predictions for the $p\gamma$ and $pp$ pion-decay and muon-damped sources with the MESE flavor contours~\cite{IceCube:2025ole} and illustrative IceCube-Gen2 projections~\cite{IceCube-Gen2:2023rds}. The predicted regions show the variation obtained from the marginal $3\sigma$ ranges of the oscillation parameters in NuFIT~6.0~\cite{Esteban:2024eli}. When the DM-induced potential dominates the vacuum terms at production, the predicted flavor compositions depend markedly on the sign of $\epsilon_\mu$.

Between these limits, the flavor fractions can vary across the measured energy range as the vacuum and DM terms compete and the propagation changes in adiabaticity. Differences among AGNs in their production mechanisms or DM environments could also yield an intermediate diffuse composition. The energy dependence and adiabaticity of the adopted spike are examined in Appendices~\ref{app:energy_transition} and \ref{app:adiabaticity}. Appendix~\ref{app:additional_flavor} shows illustrative coupling-dependent trajectories and, for electron- and tau-flavor interactions, the flavor compositions obtained when the DM-induced potential dominates the vacuum terms at production.

For the $p\gamma$ muon-damped source, the central predictions for $|\epsilon_\alpha|(m_\chi/\mathrm{GeV})^{-1}\geq10^5$ lie outside the 95\% confidence-level (C.L.) MESE contour for both signs of $\epsilon_\mu$ (Fig.~\ref{fig:comparison_icecube_2}) and for $\epsilon_\tau<0$ (Fig.~\ref{fig:comparison_icecube_tau}). If AGNs with these production and DM-spike properties dominate the diffuse flux, these predictions are in tension with the MESE measurement.

\section{Conclusions}

DM spikes around AGN black holes can enhance the potential generated by coherent forward neutrino--DM scattering to a level comparable to the vacuum oscillation terms. Our complete three-flavor calculation for pion-decay and muon-damped sources finds flavor compositions at Earth that can differ substantially from vacuum predictions and survive phase averaging. The shifts depend on the strength, sign, and flavor structure of the interaction, as well as the net DM particle--antiparticle asymmetry.

Because individual AGNs do not yet have measured flavor compositions, we compared our predictions with IceCube's diffuse measurement assuming a population of AGNs with similar production conditions and DM spikes. The strong-potential $p\gamma$ muon-damped prediction lies outside the 95\% C.L. MESE contour for either sign of $\epsilon_\mu$, suggesting tension if such AGNs dominate the flux. Improved diffuse measurements and flavor-resolved observations of individual AGNs will provide more direct tests.

\acknowledgments
\section*{Acknowledgments}
We thank Garv Chauhan and Sudip Jana for useful discussions. The work of EG, AI and YP is supported by the Collaborative Research Center SFB1258 and by the Deutsche Forschungsgemeinschaft (DFG, German Research Foundation) under Germany's Excellence Strategy - EXC-2094 - 390783311. The work of BD was partly supported by the US Department of Energy under grant No.
DE-SC0017987 and by a Humboldt Fellowship from the Alexander von Humboldt Foundation. BD thanks the T30d group at TUM for local hospitality. 

\section*{Note Added}
While finalizing this work, we became aware of a concurrent work~\cite{Tseng:2026evn} discussing a similar idea.


\appendix
\renewcommand{\thesection}{\Alph{section}}

\refstepcounter{section}
\section*{Appendix \thesection: Adiabaticity through the dark matter spike}
\phantomsection
\label{app:adiabaticity}

We quantify here the adiabaticity of neutrino propagation through the DM profile introduced in Eqs.~\eqref{eq:nfw_profile} and \eqref{eq:spike_profile}. The results presented in the main text are obtained by solving the complete radial Schr\"{o}dinger equation in Eq.~\eqref{eq:radial_evolution}, including nonadiabatic transitions. The adiabaticity parameter introduced below provides a diagnostic of where such transitions become relevant.

At each radius, $U^m(r)$ diagonalizes the local Hamiltonian,
$[U^m(r)]^\dagger H_{\rm eff}(r)U^m(r)
=D_m(r)\equiv\operatorname{diag}(\lambda_1,\lambda_2,\lambda_3)$.
Writing the flavor-amplitude vector as
$\boldsymbol{\nu}_f=U^m\boldsymbol{\nu}_m$ gives
\begin{equation}
i\frac{d\boldsymbol{\nu}_m}{dr}
=
\left[
D_m-i(U^m)^\dagger\frac{dU^m}{dr}
\right]\boldsymbol{\nu}_m.
\label{eq:app_matter_basis}
\end{equation}
The second term arises from the radial variation of the instantaneous eigenstates. Its off-diagonal elements, $-i\langle i|d j/dr\rangle$ for $i\neq j$, couple different propagation eigenstates, while its diagonal elements contribute to their phases~\cite{Kuo:1989qe}. Adiabatic following requires each off-diagonal coupling to be small compared with the corresponding eigenvalue separation $|\lambda_i-\lambda_j|$. Differentiating $H_{\rm eff}|j\rangle=\lambda_j|j\rangle$ and projecting onto $\langle i|$ gives
$\langle i|dH_{\rm eff}/dr|j\rangle
=(\lambda_j-\lambda_i)\langle i|d j/dr\rangle$ for $i\neq j$.
We therefore define
\begin{align}
\gamma_{ij}(r)
&\equiv
\frac{|\lambda_i-\lambda_j|}
{\left|\langle i|d j/dr\rangle\right|}
=
\frac{|\lambda_i-\lambda_j|^2}
{\left|\langle i|dH_{\rm eff}/dr|j\rangle\right|}.
\label{eq:app_adiabaticity}
\end{align}
The minimum three-flavor adiabaticity parameter is
$\gamma_{\min}(r)\equiv\min_{i<j}\gamma_{ij}(r)$.
Adiabatic evolution is expected when $\gamma_{\min}\gg1$ along the trajectory~\cite{Mikheyev:1985zog, Dighe:1999bi, Lunardini:2000swa}.

For a potential acting only on flavor $\alpha$,
$H_{\rm eff}=H_{\rm vac}+V_\alpha(r)P_\alpha$, where
$P_\alpha=|\nu_\alpha\rangle\langle\nu_\alpha|$.
Since $dH_{\rm eff}/dr=(dV_\alpha/dr)P_\alpha$,
Eq.~\eqref{eq:app_adiabaticity} becomes
\begin{equation}
\gamma_{ij}(r)
=
\frac{|\lambda_i-\lambda_j|^2}
{|dV_\alpha/dr|\,
 |U_{\alpha i}^{m*}U_{\alpha j}^{m}|}.
\label{eq:app_single_flavor_gamma}
\end{equation}
We evaluate this expression using the complete three-flavor
Hamiltonian at each radius. For an isolated, locally linear avoided
crossing, the Landau--Zener approximation gives
$P_{ij}^{\rm hop}\simeq
\exp[-\pi\gamma_{ij}(r_c)/4]$~\cite{Landau:1932vnv,1932RSPSA.137..696Z, Parke:1986jy}.
The results in the main text instead use the full numerical evolution.

\begin{figure*}[t]
    \centering
    \includegraphics[width=\textwidth]{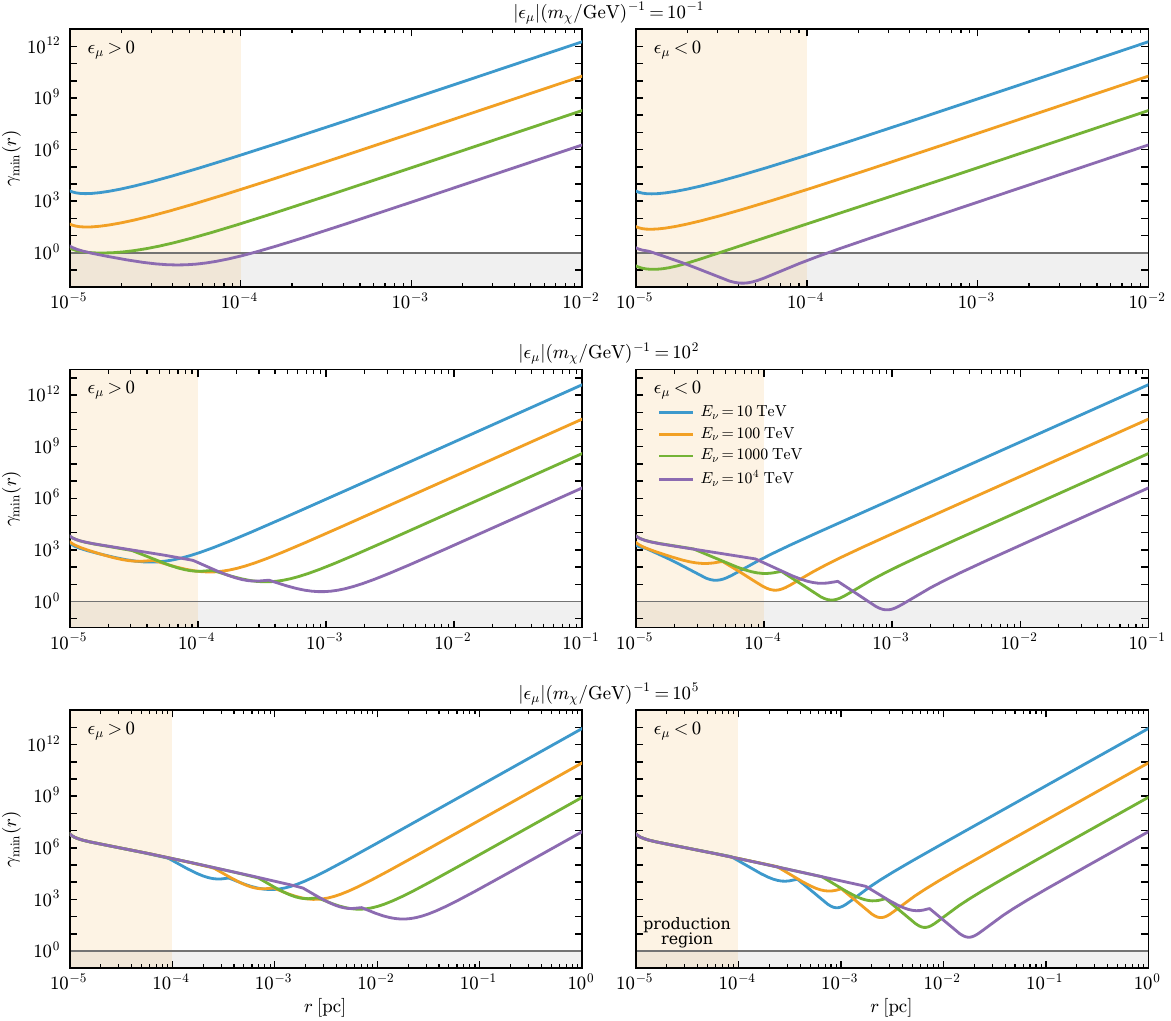}
    \caption{Minimum three-flavor adiabaticity parameter along the outward trajectory through the NGC~1068 DM spike. From top to bottom, the rows show $|\epsilon_\mu|(m_\chi/\mathrm{GeV})^{-1}=10^{-1}$, $10^2$, and $10^5$; the left and right columns show $\epsilon_\mu>0$ and $\epsilon_\mu<0$ for a particle-only DM background. Blue, orange, green, and purple denote source-frame energies $E_\nu=10$, $100$, $10^3$, and $10^4\,\mathrm{TeV}$, respectively. The shaded band marks the neutrino-production region, with propagation initiated at $r_{\rm p}=\SI{e-5}{pc}$. The horizontal line marks $\gamma_{\min}=1$. The radial range increases from the upper to the lower row to display the high-energy minima. We assume normal ordering and $\epsilon_e=\epsilon_\tau=0$.}
    \label{fig:app-adiabaticity}
\end{figure*}

Fig.~\ref{fig:app-adiabaticity} shows $\gamma_{\min}(r)$ along the outward trajectory from $r_{\rm p}=\SI{e-5}{pc}$ for normal ordering, a particle-only DM background, and a potential restricted to the muon flavor ($\epsilon_e=\epsilon_\tau=0$). The upper, middle, and lower rows correspond to $|\epsilon_\mu|(m_\chi/\mathrm{GeV})^{-1}=10^{-1}$, $10^2$, and $10^5$, respectively. The columns show positive and negative $\epsilon_\mu$, and the four curves in each panel correspond to source-frame energies $E_\nu=10,\,100,\,10^3,\,10^4\,\mathrm{TeV}$.

For $|\epsilon_\mu|(m_\chi/\mathrm{GeV})^{-1}=10^{-1}$, propagation is adiabatic at $10$--$100\,\mathrm{TeV}$, but the minima for positive and negative $\epsilon_\mu$ fall to approximately $(0.968,0.109)$ at $10^3\,\mathrm{TeV}$ and $(0.196,0.0171)$ at $10^4\,\mathrm{TeV}$. This loss of adiabaticity does not imply a large flavor shift: the resulting fractions remain close to the vacuum prediction, as shown in Fig.~\ref{fig:energy_dependence_muon_potential} for a muon-damped source.

At $|\epsilon_\mu|(m_\chi/\mathrm{GeV})^{-1}=10^2$, the corresponding minima are $(13.9,1.21)$ at $10^3\,\mathrm{TeV}$ and $(3.72,0.325)$ at $10^4\,\mathrm{TeV}$. The positive-$\epsilon_\mu$ trajectory remains more nearly adiabatic, whereas the negative-$\epsilon_\mu$ trajectory becomes nonadiabatic at high energy. For $|\epsilon_\mu|(m_\chi/\mathrm{GeV})^{-1}=10^5$, even the $10^4\,\mathrm{TeV}$ minima are $(71.9,6.28)$, so propagation is approximately adiabatic over the displayed source energies. All predictions nevertheless use the complete numerical evolution, including at higher source energies reached through cosmological redshift.

\refstepcounter{section}
\section*{Appendix \thesection: Energy dependence of DM-induced flavor conversion}
\phantomsection
\label{app:energy_transition}

The competition between the vacuum and DM-induced terms produces an energy dependence reminiscent of the solar MSW transition. The matter potential is energy independent, whereas $H_{\rm vac}\propto E^{-1}$; increasing energy therefore enhances the relative importance of the DM term. Propagation through the spike can also become nonadiabatic, as discussed in Appendix~\ref{app:adiabaticity}. Fig.~\ref{fig:energy_dependence_muon_potential} shows the resulting flavor fractions for an idealized $p\gamma$ muon-damped source. Solid curves illustrate an unredshifted source, $E_{\rm src}=E_\oplus$, while dashed curves average over a source population following the star-formation-rate density (SFRD) parametrization of Ref.~\cite{Elias-Chavez:2018dru}, with $E_{\rm src}=(1+z)E_\oplus$, where $z$ is the source redshift.

For $|\epsilon_\mu|(m_\chi/\mathrm{GeV})^{-1}=10^{-1}$, both signs remain close to the vacuum-oscillation composition $(f_e,f_\mu,f_\tau)_\oplus\simeq(0.176,0.448,0.376)$ throughout the plotted range. In the adiabatic strong-potential limit, an initial $\nu_\mu$ evolves into $\nu_3$ for $\epsilon_\mu>0$ or $\nu_1$ for $\epsilon_\mu<0$, giving the respective limiting compositions $(0.022,0.549,0.429)$ and $(0.677,0.076,0.247)$. At the intermediate benchmark of $10^2$, the positive-$\epsilon_\mu$ curves approach their strong-potential limit as energy increases. The negative-$\epsilon_\mu$ curves instead rise toward that limit and then turn back at higher energies, where nonadiabatic transitions become important. For $10^5$, both solid curves remain close to their strong-potential limits across the displayed energies. The redshift-averaged negative-$\epsilon_\mu$ curve shows a modest departure at the highest observed energies, which sample still higher energies at production.

\begin{figure*}[t]
    \centering
    \includegraphics[width=\textwidth]{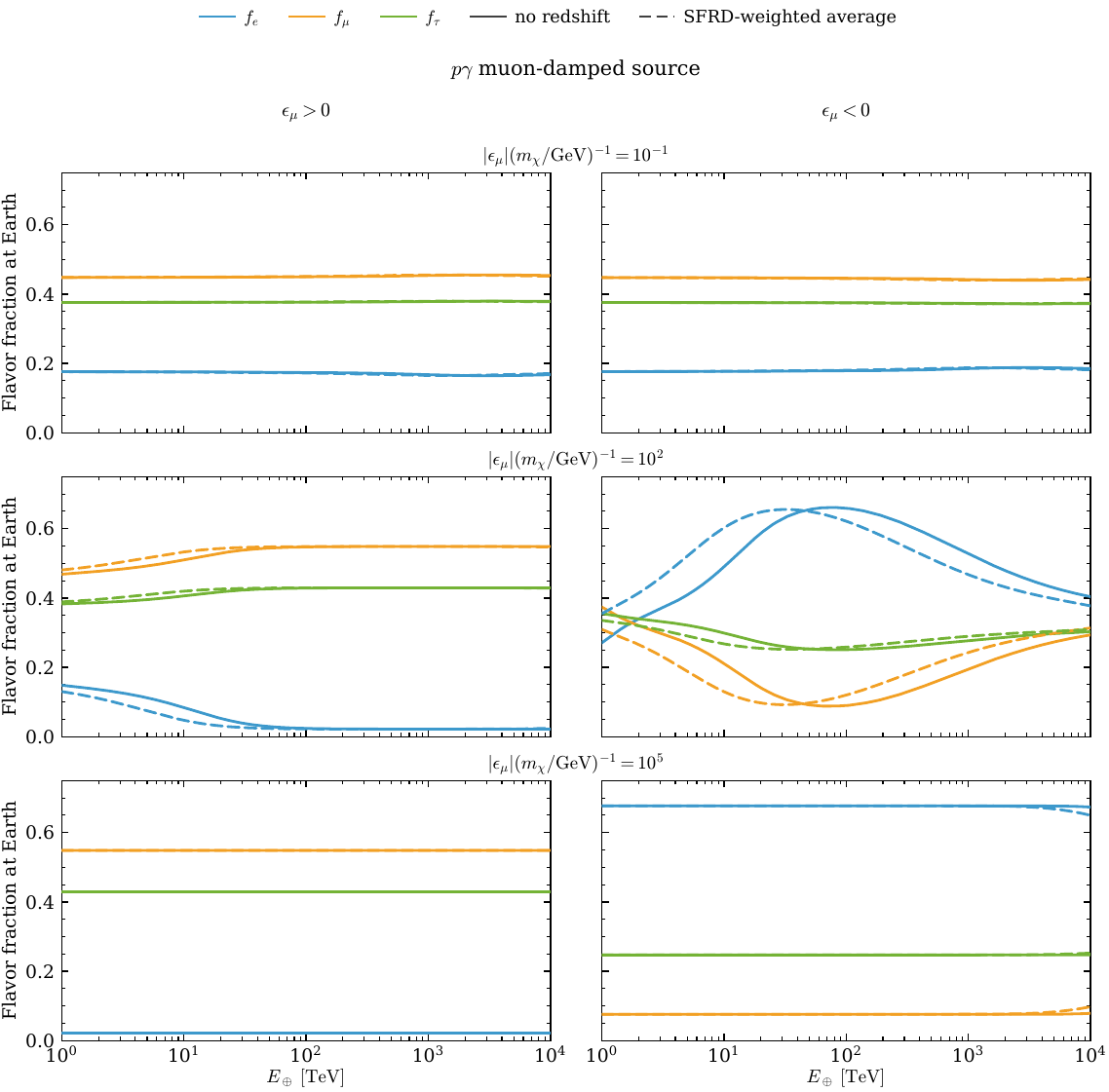}
    \caption{Energy dependence of the flavor composition at Earth for an idealized $p\gamma$ muon-damped source. The upper, middle, and lower rows correspond to $|\epsilon_\mu|(m_\chi/\mathrm{GeV})^{-1}=10^{-1}$, $10^2$, and $10^5$, respectively; the left and right columns show $\epsilon_\mu>0$ and $\epsilon_\mu<0$ for a particle-only DM background. Blue, orange, and green denote the electron, muon, and tau flavor fractions. Solid curves show an unredshifted source, while dashed curves show an SFRD-weighted source population with $E_{\rm src}=(1+z)E_\oplus$. We assume normal ordering and $\epsilon_e=\epsilon_\tau=0$, solve the complete three-flavor evolution through the spike including nonadiabatic transitions, and phase-average the subsequent propagation to Earth.}
    \label{fig:energy_dependence_muon_potential}
\end{figure*}

\refstepcounter{section}
\section*{Appendix \thesection: Coupling dependence and additional flavor interactions}
\phantomsection
\label{app:additional_flavor}

\begin{figure*}[t]
    \centering
    \includegraphics[width=\textwidth]{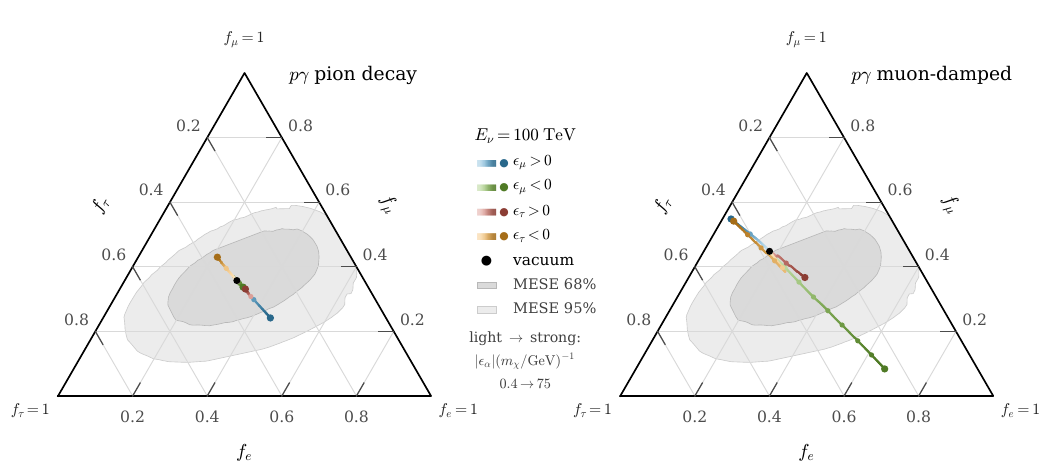}
    \caption{Flavor composition at Earth for $p\gamma$ pion-decay (left) and muon-damped (right) sources as the strength of a muon- or tau-flavor DM interaction increases. The trajectories are calculated at $E_\nu=\SI{100}{TeV}$; their colors identify the interacting flavor and the sign of its coupling, as indicated in the legend. Black markers show the vacuum-oscillation limits. Gray regions show the 68\% and 95\% MESE flavor contours.}
    \label{fig:comparison_icecube}
\end{figure*}

\begin{figure*}[t]
    \centering
    \includegraphics[width=\textwidth]{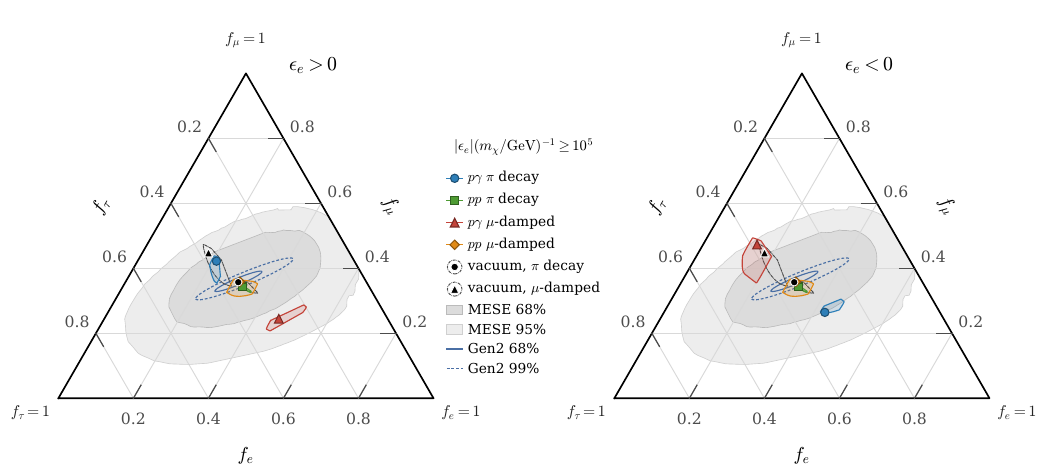}
    \caption{As in Fig.~\ref{fig:comparison_icecube_2}, but for a DM interaction acting only on the electron flavor, with $\epsilon_\mu=\epsilon_\tau=0$. The predictions are evaluated at $|\epsilon_e|(m_\chi/\mathrm{GeV})^{-1}=10^5$, representative of the limiting regime $|\epsilon_e|(m_\chi/\mathrm{GeV})^{-1}\geq10^5$, in which the DM-induced potential dominates the vacuum terms at production. The left and right panels correspond to $\epsilon_e>0$ and $\epsilon_e<0$, respectively.}
    \label{fig:comparison_icecube_electron}
\end{figure*}

\begin{figure*}[t]
    \centering
    \includegraphics[width=\textwidth]{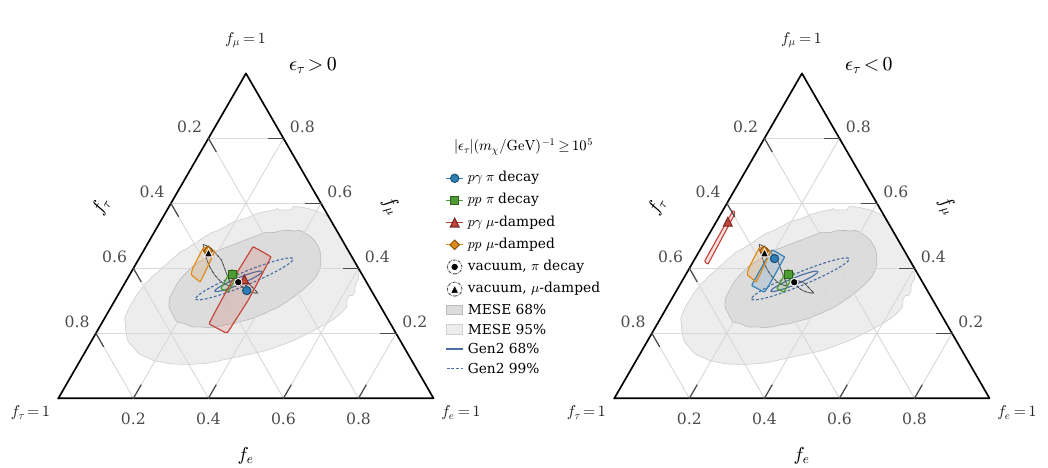}
    \caption{As in Fig.~\ref{fig:comparison_icecube_2}, but for a DM interaction acting only on the tau flavor, with $\epsilon_e=\epsilon_\mu=0$ and $|\epsilon_\tau|(m_\chi/\mathrm{GeV})^{-1}\geq10^5$. The left and right panels correspond to $\epsilon_\tau>0$ and $\epsilon_\tau<0$, respectively.}
    \label{fig:comparison_icecube_tau}
\end{figure*}

Fig.~\ref{fig:comparison_icecube} shows how the flavor compositions of $p\gamma$ pion-decay and muon-damped sources change as the DM interaction is strengthened. The direction of the shift depends on both the sign of the coupling and whether the interaction acts on the muon or tau flavor. Figs.~\ref{fig:comparison_icecube_electron} and \ref{fig:comparison_icecube_tau} extend the comparison in the main text to interactions acting on the electron and tau flavors, respectively.

\bibliography{references}

\end{document}